# The SMILE mission

G. Branduardi-Raymont[1], C. Wang[2]

[1]Mullard Space Science Laboratory, Department of Space and Climate Physics, University College London, Holmbury St Mary, Dorking, Surrey RH5 6NT, United Kingdom, e-mail: g.branduardi-raymont@ucl.ac.uk

[2]National Space Science Center, Chinese Academy of Sciences, No. 1 Nanertiao, Zhongguancun, Haidian district, Beijing 100190, China

**Abstract** This chapter describes the SMILE (Solar wind Magnetosphere Ionosphere Link Explorer) mission, currently under development as a joint project of the European Space Agency and the Chinese Academy of Sciences. SMILE aims to study the solar wind coupling with the terrestrial magnetosphere in a very novel and global way, by imaging the Earth's magnetosheath and cusps in X-rays emitted when high charge-state solar wind ions exchange charges with exospheric neutrals. SMILE combines this with simultaneous UV imaging of the northern aurora and in-situ plasma and magnetic field measurements of the magnetosheath and solar wind while flying in a highly elliptical northern polar orbit. In this chapter, after a brief introduction to the science that SMILE is targeted to investigate, the payload and the spacecraft, and progress in their development, are described. The mission operations in space, the ground segment and the scientific preparations that will ensure the optimal exploitation of SMILE measurements are also presented. The international space plasma and planetary communities are looking forward to the step change that SMILE will provide by making visible our invisible terrestrial magnetosphere.

## 1 Introduction

The solar wind coupling with the terrestrial magnetosphere is a key link in Sun-Earth interactions. Mass and energy enter geospace mainly via dayside magnetic reconnection under southward Interplanetary Magnetic Field (IMF) conditions; reconnection in the tail leads to release of energy and particle injection deep into the magnetosphere, causing geomagnetic storms and substorms. One end product are the visual manifestations of variable auroral emissions at the Earth's poles. While the basic theory of magnetospheric circulation is well known, because the microscale has been explored by many in situ measurements, the reality of how this complex interaction takes place on a global scale, and how it evolves, is still not understood. Magneto-Hydro-Dynamic (MHD) and empirical models built to simulate and describe the situation are still in need of validation. SMILE (Solar wind Magnetosphere Ionosphere Link Explorer, Branduardi-Raymont et al. 2018, BR18 henceforth) will study the full chain of events that drive Sun-Earth relations; it will provide answers to questions such as what distinguishes the modes of interaction, the characters of reconnection, what triggers geomagnetic substorms and how storms driven by Coronal Mass Ejections (CME) arise. Essentially SMILE will help establish what drives space weather, the conditions on the Sun, in the solar wind and geospace that can affect performance and reliability of our technological systems and endanger human life and health.

SMILE offers a new approach to global monitoring of geospace by imaging the magnetosheath and cusps in soft (low energy) X-rays emitted when high charge-state solar wind ions exchange charges with exospheric neutrals. SMILE is a self-standing mission coupling X-ray imaging of the magnetosheath and polar cusps (large spatial scales in the magnetosphere) with simultaneous UV imaging of global auroral distributions (mesoscale structures in the ionosphere) and in-situ solar wind/magnetosheath plasma and magnetic field measurements. SMILE will provide scientific data on the solar wind-magnetosphere interaction continuously for long, uninterrupted periods of time from a highly elliptical northern polar orbit. SMILE is a collaborative mission, from inception through to the operational phase, between the European Space Agency (ESA) and the Chinese Academy of Sciences (CAS), currently under development and due for launch at the end of 2024. This chapter presents the ongoing hardware developments and scientific preparations, and the current status, of the SMILE mission, spacecraft and payload. The chapter in Section VII of this Handbook titled 'SMILE: A novel way to explore solar-terrestrial relationships' fully focuses on the novel science that SMILE will deliver.

## 2 How we got to SMILE: SMILE precursor missions

The process of charge exchange has been known and studied since the beginning of atomic physics but only relatively recently has been recognized as being important in the astrophysics context, and more. Briefly, solar wind charge exchange (SWCX) involves highly charged solar wind ions interacting with neutral hydrogen atoms in the Earth's magnetosheath, acquiring an electron, and decaying from the excited states they are left in by emitting X-ray photons, with energies characteristic of the ions involved. The SWCX emissivity is proportional to the densities and relative velocities of the ions and neutrals, thus is highest in the Earth's magnetosheath and magnetospheric cusps. SWCX was established as being responsible for the, initially surprising, bright X-ray emission of comets (Cravens 1997). Residual X-rays from the dark side of the Moon (Schmitt et al. 1991), 'Long Term Enhancements' in the soft X-ray background observed by ROSAT (Snowden et al. 1995) and variable X-ray background in astronomical X-ray observations with lines of sight crossing the magnetosheath and the cusps (Carter et al. 2010, 2011), eventually were all attributed to the SWCX process in the Earth's magnetosheath and magnetospheric cusps under variable solar wind conditions. For more details on this process see the chapter titled 'Earth's exospheric X-ray emissions' by Jenny Carter in Section VII of this Handbook.

In practice X-rays from the Earth's magnetosphere are unwanted noise for X-ray astronomical observations 'looking out' from Earth. However, eventually they were recognised as a possibly interesting and valuable signal by space plasma and planetary scientists. Hence some in the magnetospheric community began to plan ways to exploit this signal. Missions incorporating X-ray imagers aiming to monitor the dayside magnetosheath and the cusps were proposed to NASA, e.g. MagEX (Collier et al. 2009) and STORM (Sibeck et al. 2011), and to ESA (AXIOM and AXIOM-C, Branduardi-Raymont et al. 2012), initially without success. The concept had matured sufficiently by 2015 when SMILE was selected for the collaborative mission, to be joint from beginning to end, by ESA and CAS. A team of scientists and engineers from several European countries and China, with US and Canada scientific and hardware support, make up the SMILE collaboration, dedicated to design and develop spacecraft and payload as well as the ground segment, plan orbital operations and prepare scientifically for the flight data exploitation.

## 3 Scientific payload

As mentioned in the introduction, and depicted in Fig. 1, SMILE carries a payload dedicated to the investigation of the dynamic response of the terrestrial magnetosphere to the impact of the variable solar wind, in a manner that is global and has never been tried before. Consider the impact of a solar CME arriving at the Earth. It will compress the magnetosphere and for favourable conditions of solar wind and terrestrial magnetic field orientations solar wind plasma will penetrate in the magnetosheath following magnetic reconnection. It is now well known that X-rays are produced in the dayside magnetosheath and the magnetospheric cusps by the process of SWCX. Monitoring the flux and morphological distribution of the X-rays can inform us of the locations of the bow shock, the dayside magnetopause and the cusps, and of how they change under the buffeting of the solar wind. The ultimate consequences of the solar wind penetration process are geomagnetic storms, with particles propagating to the magnetotail, undergoing reconnection again and being energised, returning to the dayside and precipitating into the polar regions, where the aurorae are the visible footprints of the whole interaction.

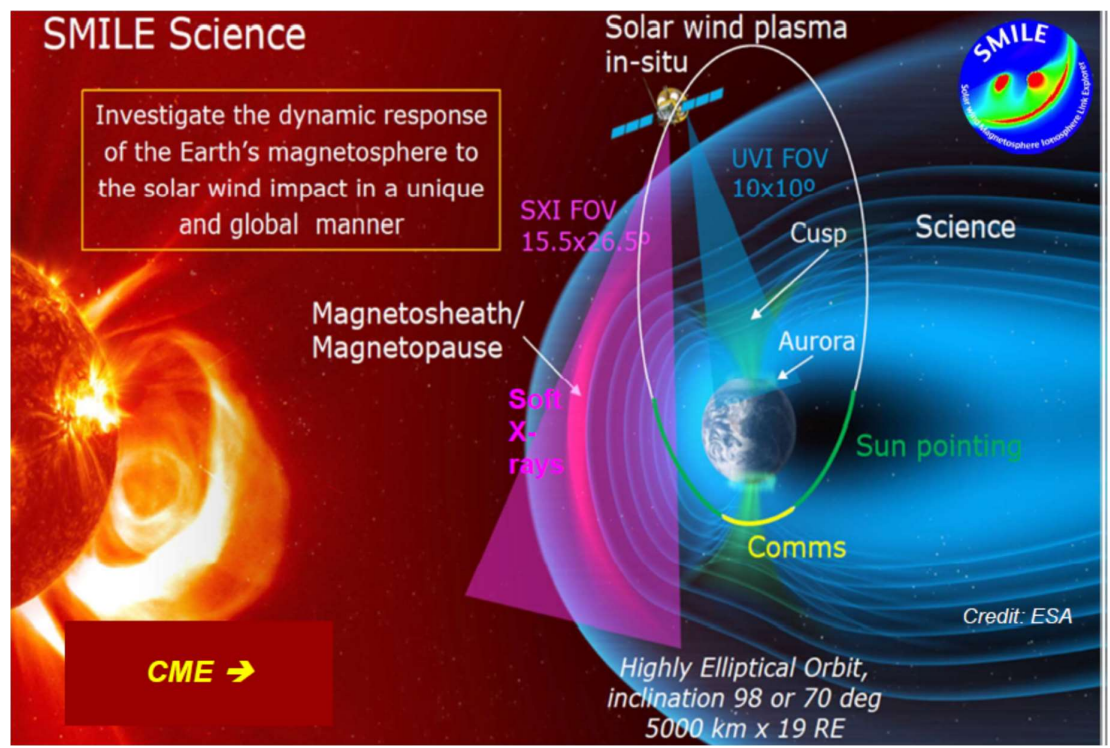

Fig. 1 The global approach of SMILE to investigating solar-terrestrial interactions. Depicted is the case of a coronal mass ejection (CME – shown as the orange arc on the left) travelling towards Earth and interacting with geospace (see text for details – Credit ESA)

So SMILE combines global X-ray imaging of the dayside magnetosheath and the cusps by the Soft X-ray Imager (SXI) with simultaneous UV imaging of the northern aurora (by the UV Imager, UVI) and in situ monitoring of the solar wind and magnetosheath conditions by the Light Ion Analyser (LIA) and the magnetometer (MAG). The principal characteristics of the payload instruments are described below. More details can be found in the ESA 'Red Book' (SMILE definition study report, BR18), as well as at the ESA and CAS websites (https://sci.esa.int/web/smile/home and http://english.cssar.cas.cn/smile/).

3.1 The Soft X-ray Imager (SXI)

The SXI is developed by the University of Leicester (Principal Investigator: Dr Steve Sembay), with hardware contributions from other institutes in the UK, from Austria, Hungary, Spain, Switzerland, Norway, and scientific support by China and the United States. SXI will observe in soft X-rays the location, shape, and motion of dayside magnetospheric boundaries, including the bow shock, the magnetopause and the cusp regions. SXI will also measure the time-dependent solar wind composition from the energy and intensity of the SWCX X-ray emission lines.

Fig. 2 shows CAD drawings of (left) the SXI telescope, with the Front End Electronics (FEE) and other subsystems highlighted, and (right) the compact X-ray camera which constitutes the middle-lower section of the telescope.

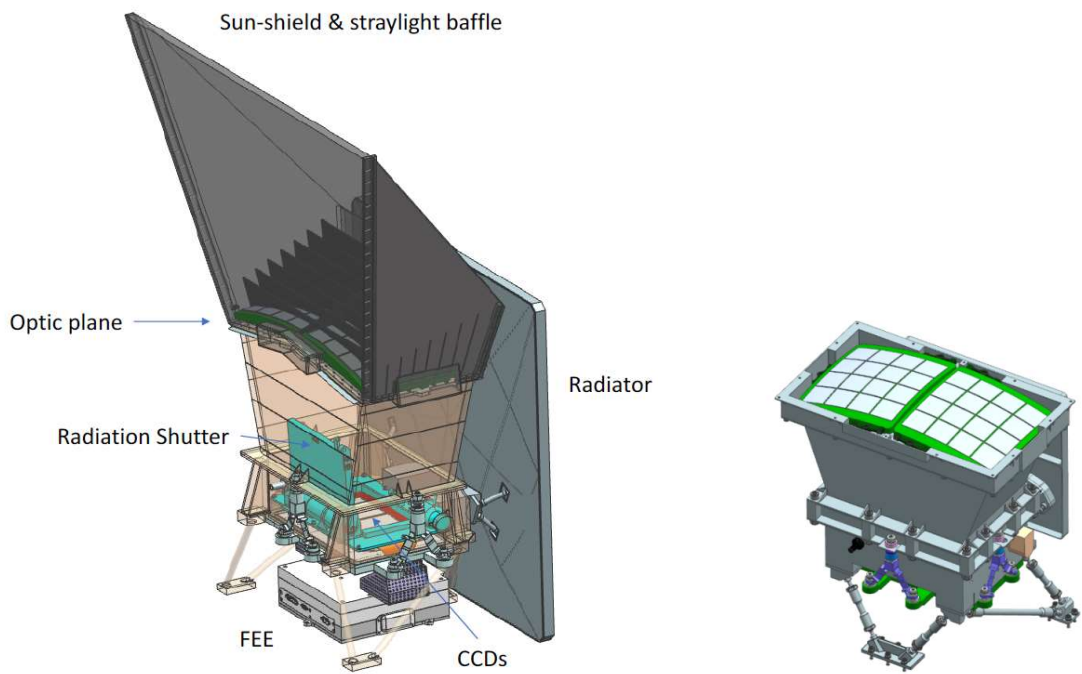

Fig. 2: CAD drawings of (left) the SXI telescope and its main subsystems, and (right) the X-ray camera showing the supporting grid structure of the micropore optic arrays. From BR18.

This wide field of view (FOV) telescope focuses soft X-rays at grazing incidence by means of a light-weight, lobster-eye-type silicon micropore optic (MPO). At the focal plane are two large format Charge Coupled Devices (CCDs, manufactured by Teledyne-e2v) where the X-rays are detected. The CCDs operate over the primary science energy range 0.2 – 2.5 keV; their operational temperature of -100 C is achieved by the use of a passive radiator (see Fig. 2) to which the CCDs are connected by thermal straps. A large optical baffle, with internal baffle vanes, is mounted above the camera to avoid solar and bright Earth straylight contamination, and an optical filter is applied on the MPOs. The main parameters of the SXI (from BR18) are listed in Table 1. CCD operations and initial X-ray signal processing are controlled by the FEE located under the focal plane, and overall instrument control and collation/compression of the data take place in the Data Processing Unit (DPU) positioned away from the instrument on the PayLoad Module (PLM: this is the plate on which SXI and UVI are mounted, which interfaces with the spacecraft PlatForm, PF – see section 4.1). A shutter, mounted above the focal plane, will be closed as the spacecraft approaches perigee to shield the CCDs from the high levels of radiation associated with the trapped particle belt near the Earth.

| Parameter | |
|---|---|
| Field of view (FOV) | 16° x 27° |
| Focal length | 30 cm |
| Point Spread Function FWHM | ~ 8 arcmin |
| Optic effective area | 15 cm$^2$ |
| Primary science operational range | 0.2 – 2.5 keV |
| CCD Quantum Efficiency | 0.89 |
| CCD energy resolution FWHM | 50 eV at 500 eV |
| CCD frame integration time | ~ 4 sec |
| Instrument mass | 36 kg |
| Optic mass | < 1 kg |

Table 1 Main parameters of the Soft X-ray Imager

During most of the SMILE orbit the SXI CCDs operate in photon counting mode, with x-y position on the CCD, energy and time stamp being returned for every detected X-ray photon, classified as such in the FEE. The CCD frame readout time is of the order of a few seconds, to avoid photon pile-up. The data are downloaded to Earth where the main image processing (e.g. background subtraction, vignetting correction, binning of the images) is performed.

Examples of scientific requirements on the SXI for relatively strong solar wind conditions can be summarised as: 1) For solar wind flux > 4.9 x10$^8$ cm$^{-2}$ s$^{-1}$ and from positions in orbit greater than 15 Earth radii (R$_E$) geocentric, the location of the subsolar magnetopause shall be determined with an accuracy better than 0.5 R$_E$ and better than 5 min time resolution, and also 2) the poleward and/or equatorward edges of the mid-altitude cusp shall be determined with a spatial resolution of at least 0.25 R$_E$ and a time resolution of at least 5 min.

The ability of the SXI to meet such science requirements, and the consequent instrument requirements, is verified by simulation. This depends not only on calculations of the expected signal strength of the primary science target (solar wind charge exchange emission within the magnetosheath and cusps) but also has to include the expected background signal from the astrophysical soft X-ray and the particle induced background. Simulations use MHD codes and input solar wind parameters (observed or artificial) to predict in 3 dimensions (3D) the density and velocity of solar wind ions once they have penetrated into the magnetosheath; these 3D 'cubes' are then integrated along the line of sight that SMILE SXI will have through the magnetosheath from different parts of the orbit; the result is combined with the neutrals density (modelled) and the SWCX emissivity is calculated; as the final step, the emissivity is input to the SXI instrument simulator code to produce the expected count images. An example of simulated images is provided in Fig. 3.

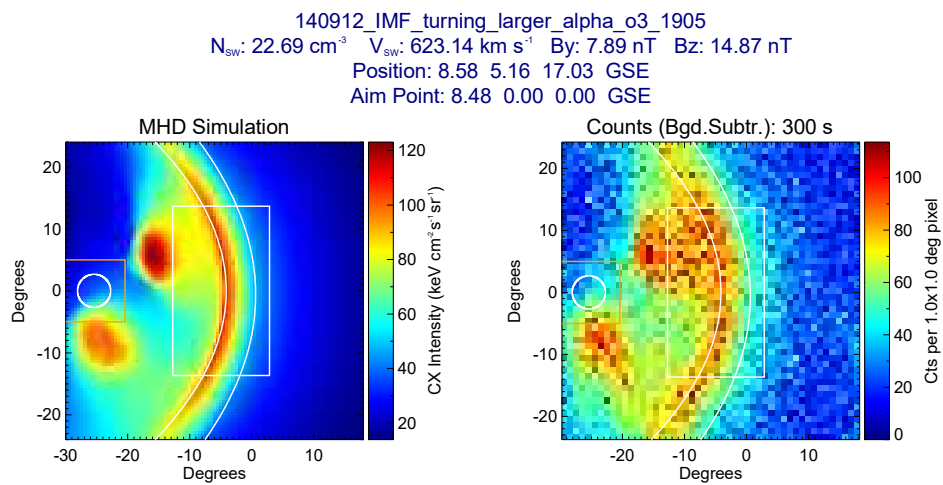

Fig. 3: Left – X-ray emissivity map produced from MHD simulations using parameters measured during a solar storm on 12 September 2014 following the arrival of a CME at the Earth. Right – Simulated soft X-ray image generated by passing the emissivity map shown on the left through the SXI simulator code. The white rectangle and circle show the SXI and UVI FOVs respectively. The curved white lines indicate the locations of the (inner) magnetopause and (outer) bow shock. Credit: S. Sembay.

In particular, to demonstrate that the SXI meets the science requirements, simulations are carried out in order to establish the accuracy with which the location of the magnetopause boundary can be established. Different methods have been explored for extracting this so called 'stand-off distance' from the images (e.g. tangential direction approach, Collier and Connor 2018; boundary fitting approach, Jorgensen et al. 2019; tangent fitting approach, Sun et al. 2020; maximum flux and maximum gradient of integrated X-ray emission, Samsonov et al. 2022, see Fig. 4). By comparing the simulation results with the original input location of the magnetopause an uncertainty of up to 0.4 $R_E$ has been derived.

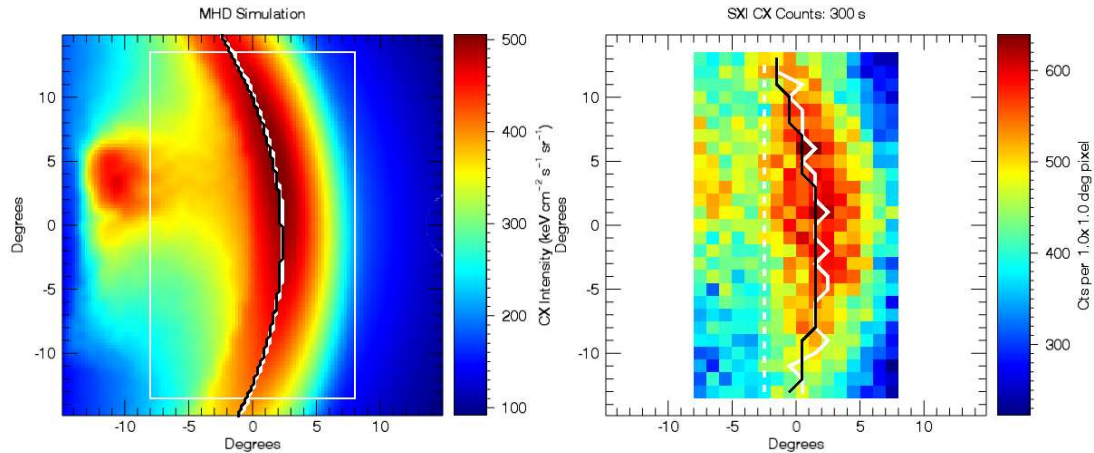

Fig. 4: Simulation results for the 16-17 June 2012 solar event with extremely high solar wind density and mostly northward IMF. Left – X-ray emissivity map for a 5 min interval on 16 June, 22:25, when the magnetospheric compression was strongest. Right – SXI count image for the emissivity shown in the left panel. The thick white lines mark the maximum of emissivity/counts (averaged over 5 pixels in the horizontal direction), the black lines are polynomial fits to the white lines and the dashed white line indicates the maximum of the averaged counts gradient. From Samsonov et al. 2022.

When the SMILE spacecraft approaches perigee SXI will have the opportunity to target the plasmasphere, which emits strongly in the EUV band. The SXI has some sensitivity to this radiation and will be able to produce images of the region by means of an operational mode that accumulates EUV images over selectable time intervals before downloading to the ground station.

3.2 The UltraViolet Imager (UVI)

Simultaneously with the SXI observations the UVI instrument will monitor Earth's northern aurora to link the processes of solar wind injection into the magnetosphere with those acting on the charged particles precipitating into the cusps and eventually producing the aurora. The UVI will employ an innovative telescope design and new filter technologies to obtain UV images of the aurora even in sunlit conditions. UVI is led by CAS with significant scientific contribution by the University of Calgary in Canada and expertise in filter coating design and manufacture from the Centre Spatial de Liège in Belgium. Prof. Xiaoxin Zhang of the National Observatory for Space Weather is Principal Investigator for the UVI.

UVI is a four mirror reflective UV imager with a 10º x 10º FOV that will encompass the whole northern aurora throughout the orbital operational phases, at high resolution spatially (150 km or better resolution from on orbit locations greater than 15 $R_E$ geocentric) and temporally (150 Rayleigh or better sensitivity over 1 min exposures). The UV bandpass (160-180 nm) is achieved by technologically advanced coating of the optical and detector surfaces which achieves >$10^{16}$ rejection in the visible. The detector is an image intensifier comprising a photocathode, microchannel plate amplifier, a phosphor and a CCD sensor. A CAD drawing of the instrument is shown in Fig. 5 (left panel). In the case of the UVI an end-to-end simulator has been developed by the University of Calgary in order to verify the performance of the instrument and plan the flight operations. An example of the simulator output is also shown in Fig. 5 (right panel).

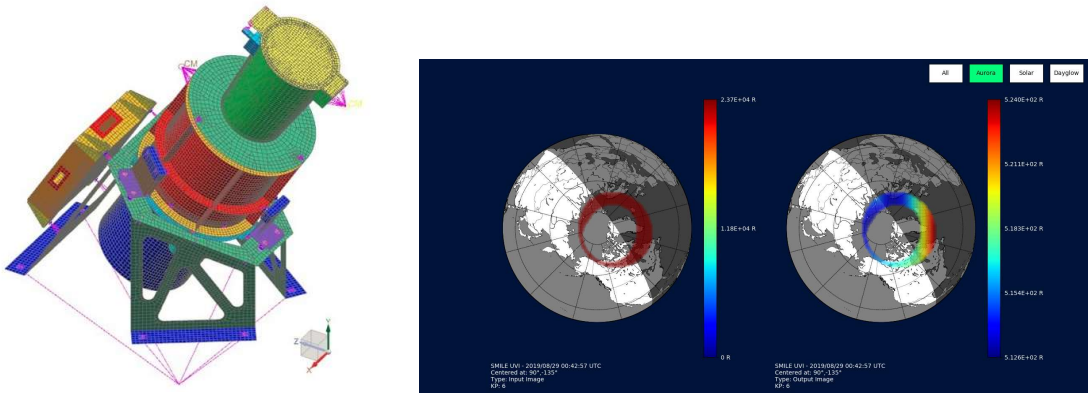

Fig. 5: Left – CAD drawing of the SMILE UV Imager. Credit: CAS. Right – Example of output from the end-to-end simulator of the UVI. Credit: University of Calgary, Canada.

3.3 The Light Ion Analyser (LIA)

The Light Ion Analyser (LIA) will measure the solar wind and magnetosheath ion distributions and determine basic moments such as density, velocity, temperature tensor, and heat flux vector. Dr Lei Dai (National Space Science Center, CAS) is the LIA Principal Investigator.

The LIA system consists of two identical instruments located on opposite sides of the CAS provided spacecraft PlatForm (PF, see section 4.1). Each of the instruments covers a solid angle of $2\pi$, such that both instruments together have a $4\pi$ FOV in azimuth. The elevation FOV for each instrument depends on the energy of the particles detected: 0 - 90° for 0.05 – 13.2 keV, 0 - 62° for 13.2 – 20 keV. Each LIA consists of an electrostatic analyser directly mounted to a back-end electronics box. A schematic representation of the LIA electron-optical system is shown in the right panel of Fig. 6 while on the left is a CAD rendition of the analyser. The LIA sensor is designed as a top-hat type electrostatic analyser with a FOV deflection system added at its entrance (marked as (2) in Fig. 6, right). Protons and $\alpha$ particles entering the analyser will be deflected in the electric field and will follow the curvature of the hemispheres (the path indicated by the red line in Fig. 6, right) to hit a position sensitive detector system (4) comprising an annular microchannel plate (MCP) with a system of 48 radial anode segments, providing directional information.

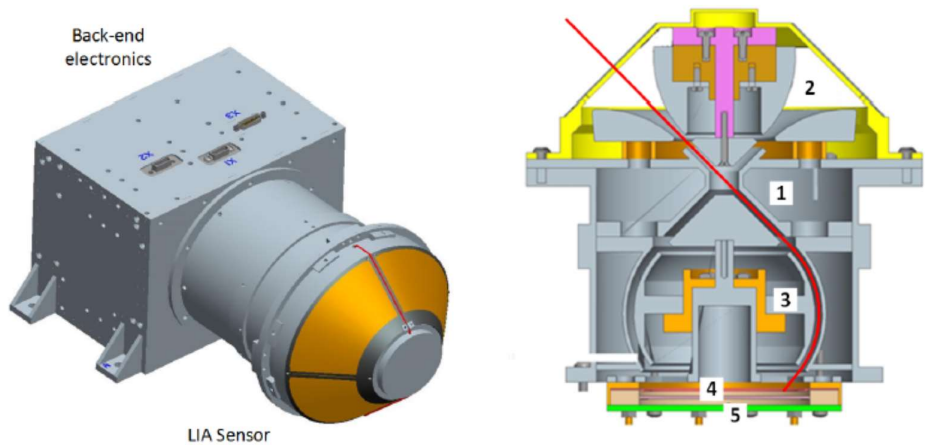

Fig. 6: Left – 3D rendition of LIA highlighting the main subsystems. Right – Cut-away view of the LIA electro-optical sensor. A typical path of a detected ion is shown as a red line. From BR18.

3.4 The Magnetometer (MAG)

The MAG instrument is developed by the CAS National Space Science Center with contributions from Austria. Dr Lei Li is MAG Principal Investigator. The scientific goal of MAG is to measure the orientation and magnitude of the solar wind and magnetosheath magnetic field. It has a dynamic range of ±12800 nT with an accuracy of 2.0 nT.

The MAG is a digital fluxgate magnetometer system consisting of two individual tri-axial sensor heads (see Fig. 7) mounted on a deployable 3 m boom (Fig. 8). The sensor heads are connected by harness to an electronics box mounted on the PayLoad Module (PLM – see sec. 4.1). The electronics unit consists of a digital processing unit, a DC/DC converter and dedicated front-end electronics for each of the magnetometer sensors.

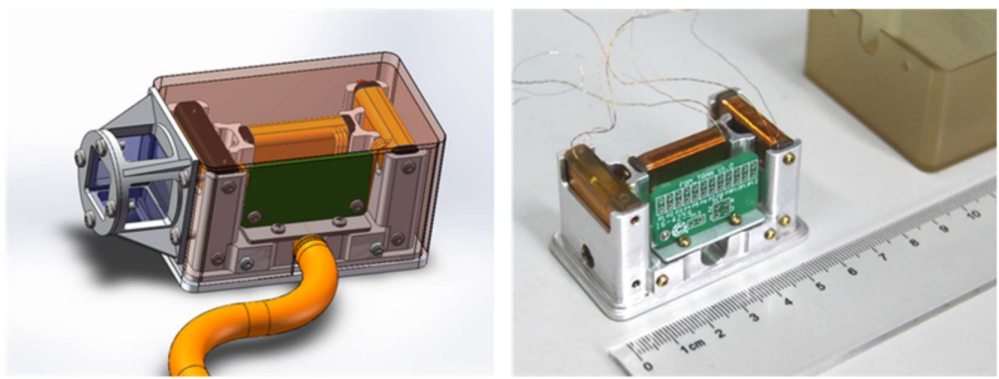

Fig. 7: The MAG sensor head. Left – 3D rendition from the CAD model. Right – Finished engineering model. From BR18.

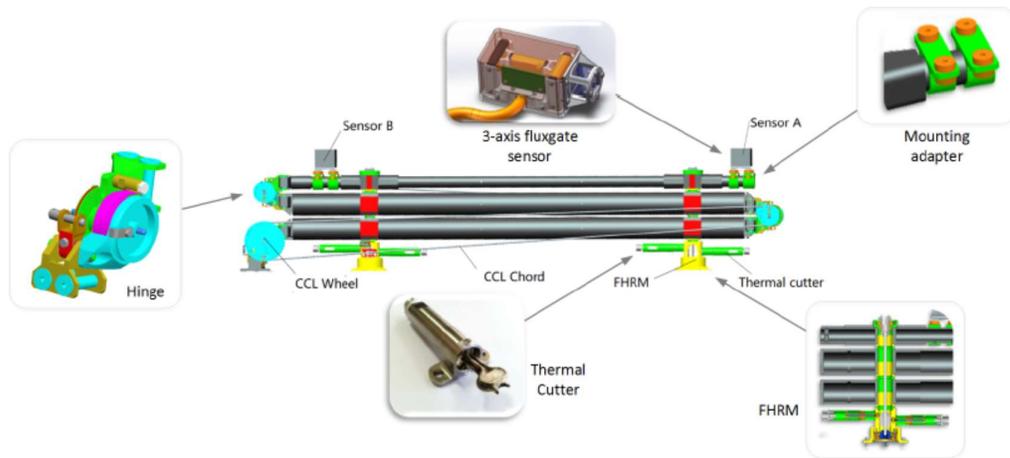

Fig. 8: The MAG deployable boom with some of its main elements highlighted. From BR18.

## 4 Spacecraft, orbit, mission design and operations

4.1 The SMILE spacecraft

The SMILE spacecraft is designed to meet the mission science and lifetime requirements: it is three-axis stabilised so that the SXI and UVI instruments can continuously point towards their magnetospheric and auroral targets. Beyond this primary driver the design of the spacecraft is driven by the division of the CAS and ESA responsibilities. This has resulted in a logical separation with respect to responsibilities and design decisions between the PlatForm (PF) and the PayLoad Module (PLM) led by CAS and ESA respectively. Fig. 9 illustrates this split of responsibilities with an image of the SMILE spacecraft in its launch configuration, also showing the locations of the instruments that make up the payload.

The PLM is acting as the top panel of the SerVice Module (SVM) in order to save mass (see Fig. 9). The X-ray and UV imagers are mounted on the PLM, as well as the folded MAG boom, to be deployed within 40 days after launch, after deployment of the spacecraft solar panels. In addition to its primary function of hosting the imagers and the MAG, the PLM provides telecommand and telemetry connections to all of them, as well as to the two LIA instruments which are mounted on the PF.

The PF is composed of the SerVice Module (SVM) and the Propulsion Module (PM). The SVM provides the power, S-band link with the ground, and the central command and control of the spacecraft including the PLM. The PM will be mainly used to raise SMILE from the initial low Earth orbit (LEO) to the final science orbit. The two LIA instruments are mounted on the PM in order to achieve a full 360º FOV in azimuth.

An artist's impression of the SMILE spacecraft in flight with deployed solar panels is given in Fig. 10.

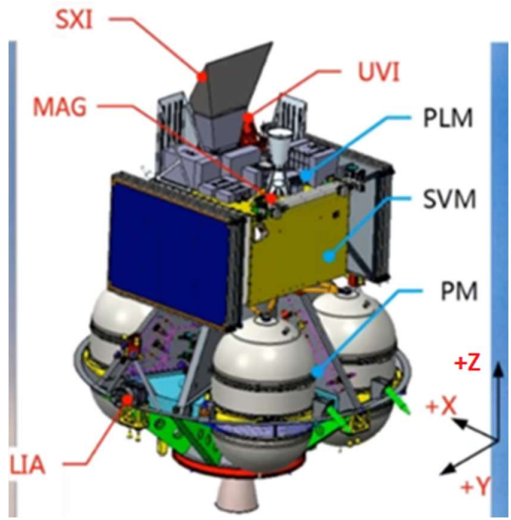

Fig. 9: The SMILE spacecraft in launch configuration.
PLM: PayLoad Module
SVM: SerVice Module
PM: Propulsion Module
SVM and PM combined make up the PlatForm (PF).
Credit: CAS/ESA.

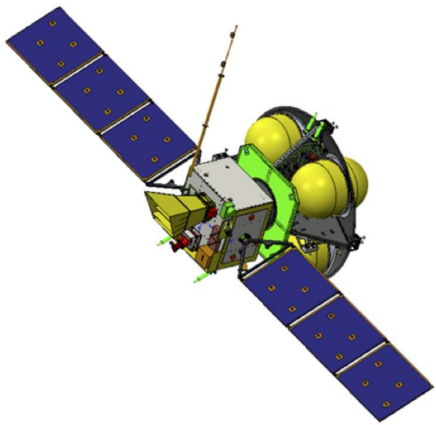

Fig. 10: Artist's impression of the SMILE spacecraft in flight.
Credit: CAS/ESA.

The propellant mass required to reach and maintain the science orbit represents two thirds of the spacecraft launch mass of 2300 kg, thus mass is a strong design driver considering the mass limitations of the Vega-C launcher (see sec. 4.3). Specific instrument and hardware constraints are also primary drivers for the mission as a whole. Examples are the radiation shielding requirements for the electronic units and instruments, which are more stringent than for a typical platform in LEO. The thermal environment required for the science operations of the SXI instrument is another key driver of the mission design. The minimum power required by SMILE is 447 W in survival mode and the maximum power of 665W is required during communications and transfer of data to the receiving ground station. The 4.2 m$^2$ solar cells can generate 850 W.

4.2 Spacecraft integration and testing

The spacecraft combines the CAS led PF and the ESA led PLM. They are designed and developed separately, tested independently, and then integrated, and tested at system level. The PF and PLM are connected through electrical and data interfaces, which are coordinated at ESA/CAS level through interface requirements and interface control documents.

At the time of writing (July 2022) Structural Thermal Models (STM) of SXI and UVI, Qualification Model (QM) of the PF and Electrical Models (EM) of LIA and MAG have been built, in Europe and China, and are being integrated in China, prior to undergoing environmental tests. Fig.s 11 to 14 show images of the instruments and the PF in their current status. Eventually the integration and testing of the SMILE Flight Model will take place at the European Space Research and Technology Centre (ESTEC) of ESA in The Netherlands, before shipping to the launch site.

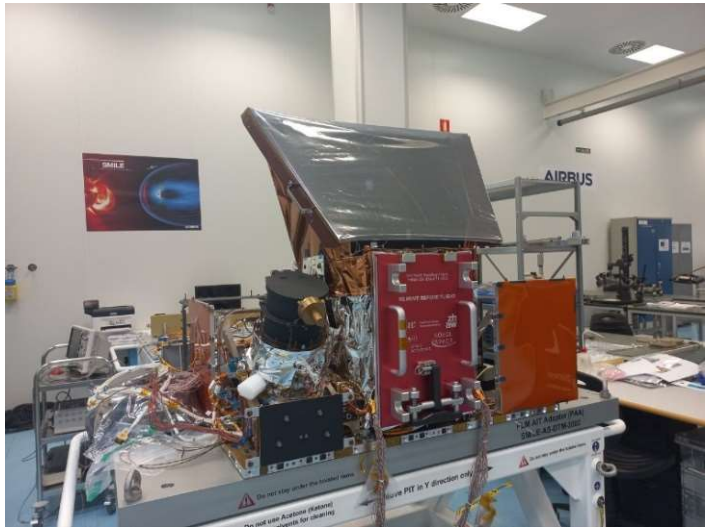

Fig. 11: STM SXI (large structure in the centre) and UVI (black cylinder on the left) mounted on the STM PLM, before shipping to China from Spain.
Credit: Airbus, Madrid, Spain.

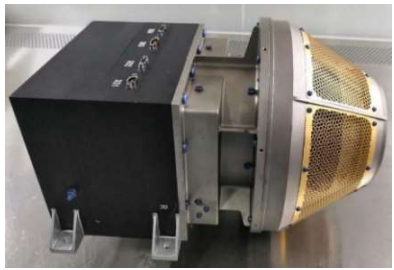

Fig. 12: EM LIA sensor and electronics.

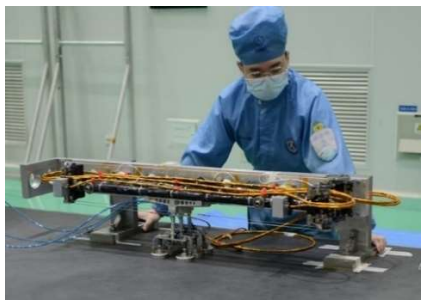
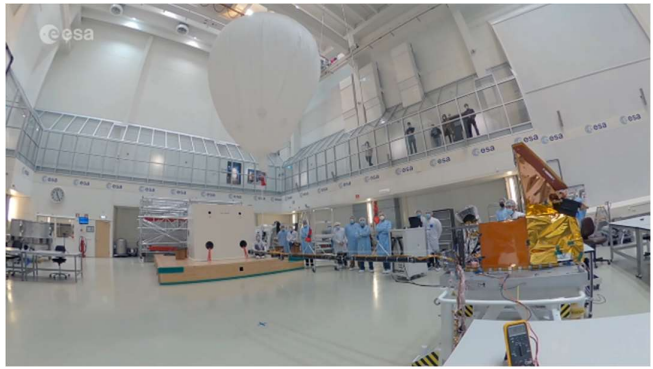

Fig. 13: MAG deployment test: Left – At instrument level, CAS/NSSC. Right – At ESA's European Space Research and Technology Centre (ESTEC), The Netherlands, after integration with the PLM.

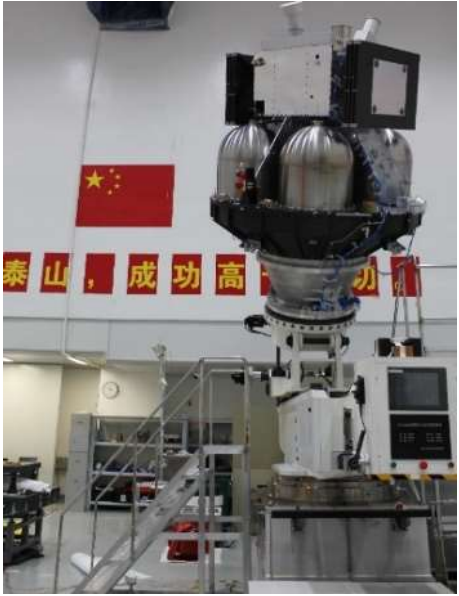

Fig. 14: SMILE QM PF undergoing vibration tests in China. At the top is the SVM with the un-deployed solar panels; at the bottom, the PM with the four propellant tanks. Credit: CAS/NSSC.

4.3 Orbit and mission design

The SMILE spacecraft will be launched from the Centre Spatial Guyanais, Kourou, French Guiana. SMILE is compatible with both a launch on Vega-C, as a single passenger, and one on Ariane 6-2 as dual launch. The down-selection of the launcher will be made after the Mission Critical Design Review, in spring 2023. If Vega-C is adopted, SMILE will be launched into a 73° inclined LEO orbit, typically 700 km circular. With Ariane 6-2 most likely it will be sent into a Sun-synchronous orbit, 700 km circular, together with an Earth observation satellite, in order to increase the likelihood of finding a primary passenger. In this case, it will be necessary to wait in LEO so that the Right Ascension of the Ascending Node of the orbit is in the required range, which could take up to 6 months. Using its own chemical propulsion the spacecraft will then transfer to its final highly elliptical science orbit, with apogee of 20 $R_E$ and perigee of 5000 km. This orbit, of 51 hours period, ensures a long observation time around apogee for monitoring the regions of interest (about 40 hours of continuous operations) with a relatively low perigee for downloading the scientific data without diving too deep into the Earth's radiation belts. A significant amount of simulations were carried out during the initial phases of designing the mission in order to establish an orbit that maximises the observing efficiency of the magnetospheric targets of interest: nose of the magnetopause, magnetospheric cusps and northern aurora. An example of the simulation output is shown in Fig.15. This uses a sophisticated orbit, visibility and FOV simulator developed by Dr Andy Read of Leicester University, in order to carry out a trade-off to select the baseline orbit and optimise the SXI visibility efficiency. At the top left of the diagram are orbit projections on three planes in the GSE (geocentric solar ecliptic) reference frame. At the bottom left is the visibility plot, in terms of orbit fractions, taking observing constraints into account. On the right is the larger FOV window displaying the main targets: nose of the magnetopause (the small brown ellipse), the cusps (the two red spheres) and the northern aurora (the black circle marked at the Earth's pole). Also indicated are magnetic field lines connecting the cusps to the Earth's surface.

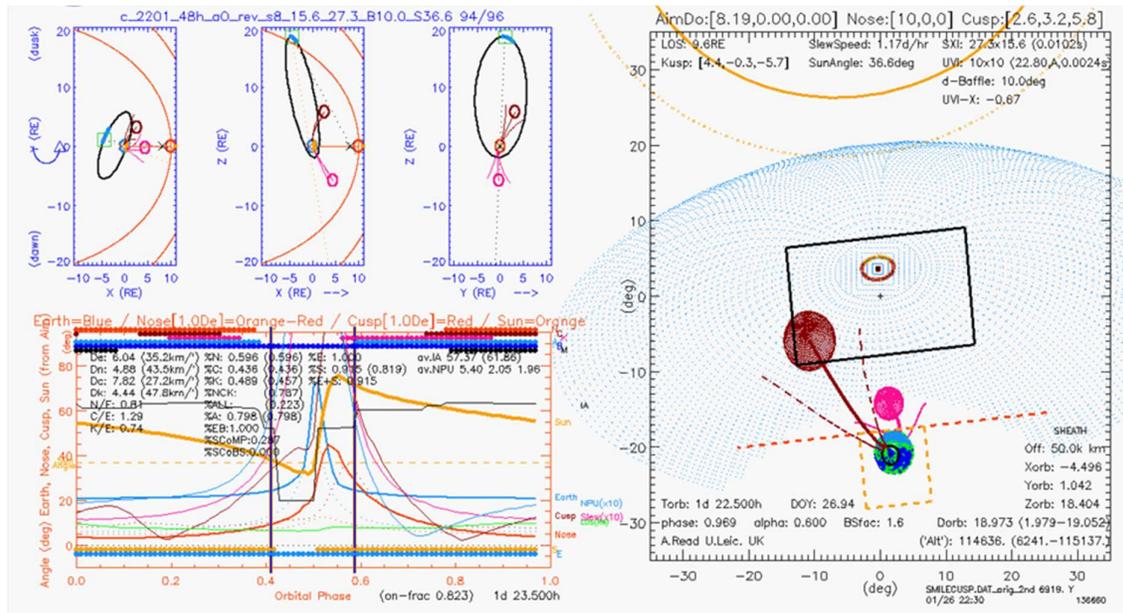

Fig. 15: Example of the output of the orbit, visibility and FOV simulator for the SMILE mission. Upper left: Orbit window. Lower left: Visibility window. On the right: FOV window. See text for details. Credit: Dr Andy Read, Leicester University.

The left panel of Fig. 16 shows how eighteen burns of the main engine will be required to take SMILE to its final highly elliptical science orbit. The diagram on the right illustrates the continuous monitoring of the magnetopause and of the northern aurora that SMILE will carry out from its orbit.

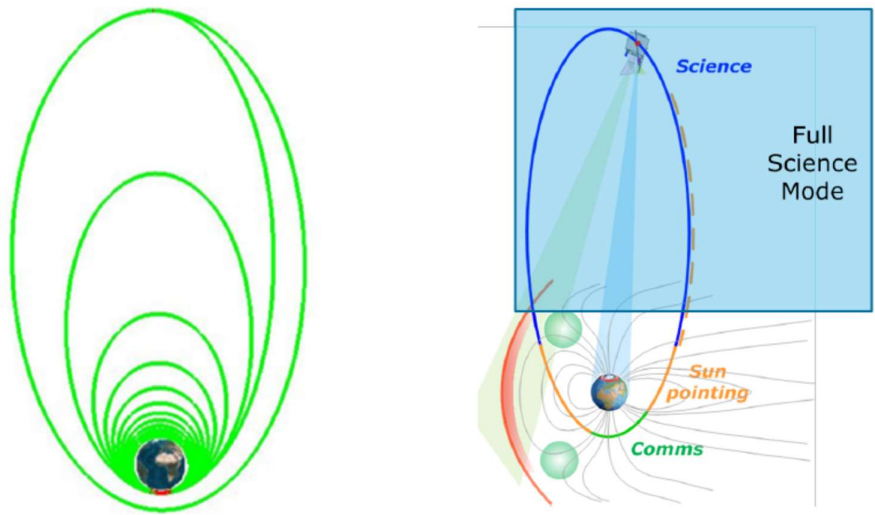

Fig. 16: Left: SMILE orbit transfer strategy. Right: SMILE operating modes along its orbit. From BR18.

4.4 Operations and ground segment

The SMILE operating modes sketched along the orbit in Fig. 16 have been defined in greater detail by taking into account the observing constraints of the imagers. These are shown in Fig. 17 and are defined as follows:

• *Full Science Mode* (SXI, UVI, LIA, MAG collecting data) above 50000 km (SXI lower altitude operational limit) – SXI aims at the nose of the magnetopause
• *Reduced science mode 1*: (SXI, LIA, MAG collecting data) above 50000 km (SXI lower altitude operational limit) – SXI aims at the nose of the magnetopause
• *Reduced Science Mode 2*: (UVI, LIA, MAG collecting data) from 30000 km (UVI lower altitude operational limit) to 50000 km (SXI lower altitude limit) – UVI points to the aurora
• *Reduced Science Mode 3*: (LIA, MAG collecting data) from perigee altitude to 30000 km (UVI lower altitude limit) – Sun acquisition: solar panels point to Sun direction

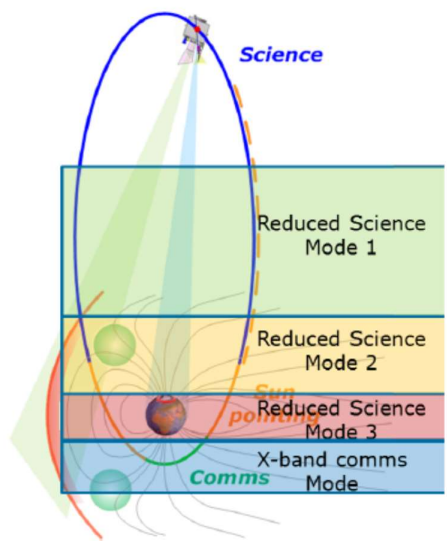

Fig. 17: SMILE operating modes at different altitudes. From BR18.

Responsibility for the SMILE ground segment and operations is shared by CAS and ESA.

The CAS Ground Support System (GSS) is composed of the Mission Centre, the Chinese Space Science Data Centre, and the CAS X-band ground stations; part of the ground system are also the Chinese Control Centre, the S-band ground stations and the Science Application System which deals with planning observations and data processing.

The ESA Ground Segment comprises ESAC (European Space Astronomy Centre near Madrid, Spain) which incorporates the Science Operations Centre (SOC) and ESDC, the ESAC Science Data Centre, as well as ESOC (European Space Operations Centre, Darmstadt, Germany), including ESA's tracking stations network (ESTRACK) connecting to the ESA's S-band and X-band ground stations.

# 5 SMILE Science Working Groups, Science Working Team and Consortium

## 5.1 SMILE Science Working Groups (SWGs)

A number of SMILE Science Working Groups (SWG) have been established in order to support the SMILE mission development and operations, ensuring that the science objectives are achieved and maximised, and adding value to SMILE science through the contributions of complementary measurements on the ground and in space. The SWGs are summarised figuratively by the cartoon in Fig. 18.

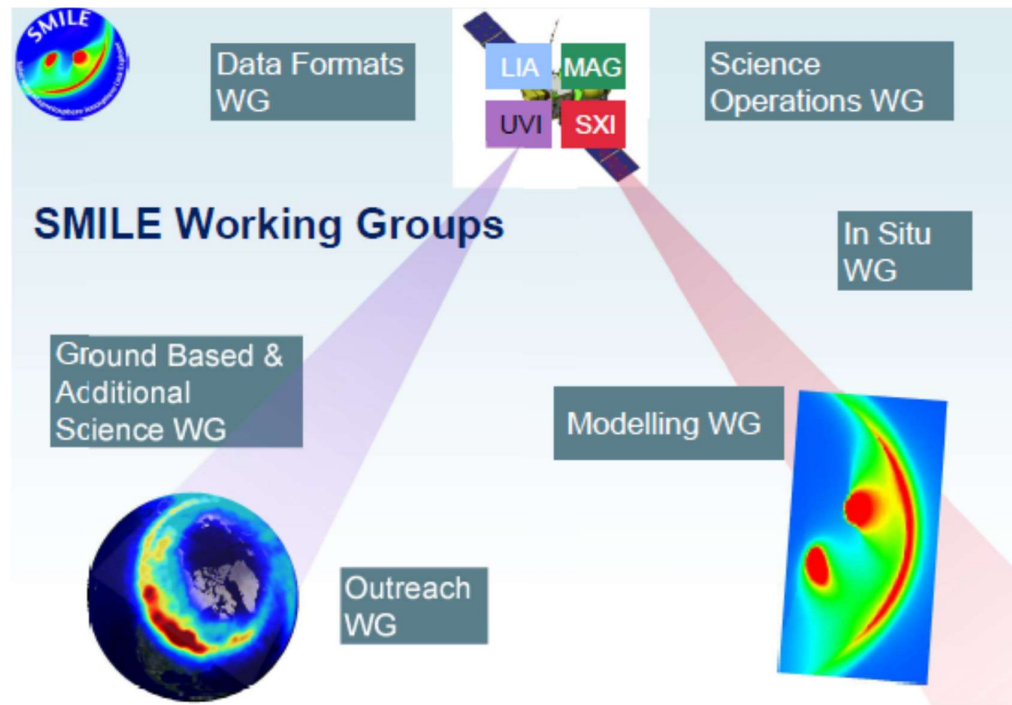

Fig. 18: Cartoon summarising the SMILE Science Working Groups.

**Science Operations WG**
This WG has been established to coordinate the preparations and the smooth running of SMILE's flight operations. Current activities comprise the preparation of the Master Science Plan, including science mode operations and calibrations, the definition of planning/commanding requirements for the spacecraft and the four instruments in the payload, and the interfaces with the ground systems of both CAS and ESA.

**In Situ WG**
The WG activity is centred on optimising the design, operations and calibrations planning of the in situ instrument package, LIA and MAG. The tasks of the WG include maintaining under review science objectives, pointing and measurement accuracy requirements, operational modes, calibration plans and data products.

**Data Formats WG**
Studies and discussions are ongoing in order to establish the most effective way to incorporate the remote sensing and in situ data which SMILE will return into a user-friendly framework, taking into account the different formats used. The WG has picked up momentum from the monthly teleconferences where topics such as the science data exchange process, commonly used formats of remote sensing and in situ data, and the conversion between them, data archiving and version control, are presented and discussed in view of their eventual implementation.

**Ground-Based and Additional Science WG**
The remit of this WG is to coordinate and later implement joint observing campaigns between SMILE and a range of solar-terrestrial space physics projects and experiments at a variety of geographical locations around the globe, whether they be from space-based or ground-based instrumentation. SMILE will maximise the best science return by eliciting support of other experiments at a range of spatial and temporal scales. Links with ground observers are already well established and contacts with operational and forthcoming space missions are taking place in order to coordinate observing and software efforts.

**Outreach WG**
SMILE will give important contributions to outreach and public engagement: by producing a very visual output from its observations, that is images and movies of the dynamic response of the Earth's magnetic field to the vagaries of the solar wind, it will capture the interest of the public, young and old, for a science, that of magnetic fields and particles in geospace, that would be hard to grasp without imagery. Many discussions and activities are already taking place generating contacts with organisations promoting science in primary and secondary schools, holding workshops and advocating careers in science, focusing on SMILE as a practical example of how space projects are developed, and encouraging pupils to follow its progress to launch and beyond. The WG is happy to share presentations on SMILE and outreach material and welcomes their use targeted to engage more cohorts of students.

**Modelling WG**
This WG aims to coordinate and drive the simulation and modelling activities ongoing in many SMILE collaborating institutes and leading to predict the soft X-ray images which SXI will generate. This work, simulating the changes which can be expected in magnetospheric boundary locations under differing solar wind conditions, coupled with investigations of how to extract magnetospheric boundary and cusp positions, is essential in order to optimise the SXI observing strategy once SMILE is operational. The WG is very active, with monthly teleconferences where research work is reported and discussed in detail. New participants, interested in hearing about the activities and in presenting research relevant to SMILE science, are always welcome.

More details about these SWGs, including useful links, can be found at the SMILE website https://www.mssl.ucl.ac.uk/SMILE/

5.2 SMILE Science Working Team (SWT) and Consortium

A SMILE Science Working Team (SWT), comprising the CAS Chief Scientist, the ESA Project Scientist, instruments Principal Investigators (PIs) and mission Co-PIs, has been established after the mission was formally adopted for implementation by ESA in March 2019. The SWT replaces the Science Study Team that was active during the study phase of the mission, before adoption. The SWT and the wider SMILE Consortium of scientists and technologists from all over the world meet approximately every six months to examine progress in the mission development, agree on the way ahead and review the preparations for science exploitation. All the SWGs provide reports on their activities, and Consortium members offer presentations on science and engineering topics of relevance to SMILE. The meetings alternate between a location in China and one in the West. Due to the Covid pandemic, for two and a half years these meetings have turned to be run as teleconferences, with the hope to return to face to face encounters in the not too distant future.

5.3 Data policy

The ESA and CAS science data archives will be the repository of all mission products. All scientists and the public over the world, defined as the 'users', will have access to the SMILE data deposited in the archives after a proprietary period. This proprietary period is one year after data have been received by the SWT. A PI could open the data of his/her instrument to the users during the proprietary period after approval by the SWT. Quick-look plots will be publicly available as soon as feasible and should be used for event browsing, but should not be used for publication, except with specific agreement from the SWT. Users are recommended to contact the SMILE SWT early in an analysis project to consult on the appropriate use of and publication of instrument data. Users should acknowledge the official sources of data used in all publications, presentations, and reports.

6   **SMILE impact and legacy**

The variable SWCX X-ray emissions in the magnetosheath and cusps play the role of 'unwanted background' for X-ray astrophysical observatories in LEO when their line of sight crosses these regions. With its novel approach the SMILE SXI turns this unwanted noise into a visual diagnostic tool of Sun-Earth relationships. As a self-standing mission, SMILE will provide direct scientific input to the studies of space weather by providing the remote sensing measurements needed to validate global models of solar wind-magnetosphere interactions.

SMILE has also significant potential for outreach, as outlined in sec. 5.1 by the paragraph on the dedicated WG: the images and movies that it will provide will captivate the public to science (that of the Earth's magnetic field) so far invisible, making it visible for the first time.

The cooperation with China is also a unique aspect of the SMILE mission, given this is the first time that the two agencies, ESA and CAS, are developing a joint mission from beginning to end: SMILE is a showcase of how good collaboration by scientists and engineers can overcome a variety of hurdles in order to reach a common goal.

# 8 Conclusions

SMILE will make great strides in leading us to understanding large scale phenomena in the Earth's magnetosphere by making coupling of the solar wind with magnetosphere and ionosphere visible in a global way. SMILE will help discern the modes of the magnetospheric-ionospheric interaction under a variety of interplanetary and solar wind conditions. The way in which SMILE will carry this out, by imaging the magnetosheath in SWCX X-rays, is now generally accepted as a novel and efficient approach to monitoring and investigating further solar-terrestrial interactions. SMILE will demonstrate the feasibility and power of this approach, paving the way for future more ambitious missions, carrying instrumentation e.g. to the Moon, a splendid vantage point on the global geospace, or further afield, to apply what learnt at Earth to more distant worlds in the solar system, like the outer planets.